# Scalable Memdiodes Exhibiting Rectification and Hysteresis for Neuromorphic Computing

Joshua C. Shank[1], M. Brooks Tellekamp[1], Matthew J. Wahila[2], Sebastian Howard[2], Alex S. Weidenbach[1], Bill Zivasatienraj[1], Louis F. J. Piper[2], and W. Alan Doolittle[1]*

[1] Department of Electrical and Computer Engineering, Georgia Institute of Technology, Atlanta, GA 30332, United States of America
[2] Department of Physics, Applied Physics and Astronomy, Binghamton University, Binghamton, NY 13902, United States of America

*alan.doolittle@ece.gatech.edu

Metal-Nb$_2$O$_{5-x}$-metal memdiodes exhibiting rectification, hysteresis, and capacitance are demonstrated for applications in neuromorphic circuitry. These devices do not require any post-fabrication treatments such as filament creation by electroforming that would impede circuit scalability. Instead these devices operate due to Poole-Frenkel defect controlled transport where the high defect density is inherent to the Nb$_2$O$_{5-x}$ deposition rather than post-fabrication treatments. Temperature dependent measurements reveal that the dominant trap energy is 0.22 eV suggesting it results from the oxygen deficiencies in the amorphous Nb$_2$O$_{5-x}$. Rectification occurs due to a transition from thermionic emission to tunneling current and is present even in thick devices (> 100 nm) due to charge trapping which controls the tunneling distance. The turn-on voltage is linearly proportional to the Schottky barrier height and, in contrast to traditional metal-insulator-metal diodes, is logarithmically proportional to the device thickness. Hysteresis in the *I-V* curve occurs due to the current limited filling of traps.

**Introduction**

In 1962, a neuristor circuit was theorized for propagating electrical signals without attenuation as is observed between biological neurons[1]. In 2013, a NbO$_2$ memristor was demonstrated that replicates the voltage triggered switching behavior observed in biological ion channels enabling a physical neuristor circuit to be built[2]. These NbO$_2$ memristors successfully replicate the rectification observed in voltage-gated biological ion channels[3] and the hysteresis necessary for the proposed neuristor circuit to operate[2]. The NbO$_2$ memristors are manufactured by electroforming an insulating layer of Nb$_2$O$_5$ into the semiconductor NbO$_2$, a process that introduces challenges for scaling to dense circuits. Additionally, previous work has not reported a capacitance in NbO$_2$ memristors requiring an additional parallel capacitance for the neuristor circuit to operate[2]. Herein, an Nb$_2$O$_{5-x}$ based metal-insulator-metal (MIM) device is reported that does not require electroforming, possesses a dispersive capacitance advantageous for neuromorphic circuits, and exhibits the rectification and hysteresis necessary to replicate ion channel functionality for the neuristor circuit.

The Nb$_2$O$_{5-x}$ devices function in the same way as the previously published voltage threshold switches consisting of electroformed NbO$_x$ filaments within Nb$_2$O$_5$[2]. When used in a neuristor circuit as described in *Pickett et. al.*[2] the switching device, whether it is a memdiode or the NbO$_2$ device, operates as a voltage threshold triggered switch. The neuristor oscillates when the circuit is excited by a current. Initially, a capacitor is charged in parallel with one switch, and that switch turns on when a certain threshold voltage is reached. This switch will also turn off at a voltage lower than the turn-on voltage. When this first switch turns on, the voltage increases on a second memdiode/capacitor pair which causes the second memdiode to switch. This time delayed switching raises then lowers the voltage, creating an oscillatory pulse. The memdiodes presented herein act in the same way as the prior NbO$_2$ threshold switches (both act as voltage threshold switches) with the added benefits of integrated capacitance and as-deposited switching. In this way two such memdiodes can form the basis of a neuristor by incorporating both the switching element and the capacitance into one device that does not require electroforming. At the envisioned scale for a neuromorphic system, $10^4$ synapses per neuron and $10^6$ neurons/cm$^2$,[4] electroforming each individual NbO$_2$ memristor becomes unfeasible, and with the constraint upon device size capacitor/switch integration becomes highly desirable.

Niobium Pentoxide (Nb$_2$O$_5$) has been studied as an insulating layer in MIM diodes since at least the 1960's[5]. These early studies on Nb – Anodized Nb$_2$O$_5$ – Au devices found rectification[5] and a Poole-Frenkel conduction mechanism[6]. The defect state enabling Poole-Frenkel conduction was found to lie between 0.22 – 1.2 eV[5,6] below the conduction band edge and was attributed to oxygen vacancies[6]. Additionally, it was found that upon dielectric breakdown these devices exhibited bistable switching in small conducting regions, similar to the NbO$_2$ devices used in the 2013 neuristor demonstration[7]. The rectification observed in these NbO$_2$ memristors was originally attributed the to the metal-insulator transition (MIT) occurring at 1081 K. However, there is debate in the literature regarding the operating mechanism of the electroformed NbO$_2$ devices with both a Mott MIT[2] and a Joule heat driven Poole-Frenkel mechanism[8,9] suggested. Other studies of traditional transition metal oxide memristors show multiple conduction mechanisms for different resistance states, transitioning from metallic conduction at LRS to electron hopping at intermediate states and Schottky emission at HRS[10,11]. Simulations suggest that these NbO$_2$ devices rectify through a modest temperature increase of 380 K – 400 K which significantly increases the contribution of the highly non-linear Poole-Frenkel current transport[8,9]. Most recently two negative differential resistance regions have been demonstrated in current-controlled NbO$_2$ devices, indicating the presence of a lower temperature Poole-Frenkel based transition and a higher temperature MIT.

**Results**

The *I-V* characteristics for a Metal-Nb$_2$O$_{5-x}$-Metal diode exhibit rectification and hysteresis, as shown in Figure 1a. This measurement was performed using 50 mV steps and holding for 15 seconds at each voltage step. At each voltage step the current transient was measured and is shown in Figure 1b. For voltages below turn-on, the current decreases as a function of time in a nearly exponential fashion. For voltages above turn-on, the current increases with time. As shown in Figure 1c, the current transients above turn-on have not saturated to steady state after 150 seconds indicating that a comparatively slow mechanism is responsible for the current transient. The hysteresis present in the current-voltage curves may therefore be due to the long current saturation time, allowing a broad range of frequencies for neuromorphic device operation (i.e. f > 1/150 Hz) which requires hysteretic behavior. The combination of rectification and memristive hysteresis lead to the term memdiode, consistent with previous devices that exhibit these features[12]. It is also noted that due to this diode-like behavior the traditional memristor figures of merit low resistance state (LRS) and high resistance state (HRS) are not as relevant as the turn-on and turn-off voltages.

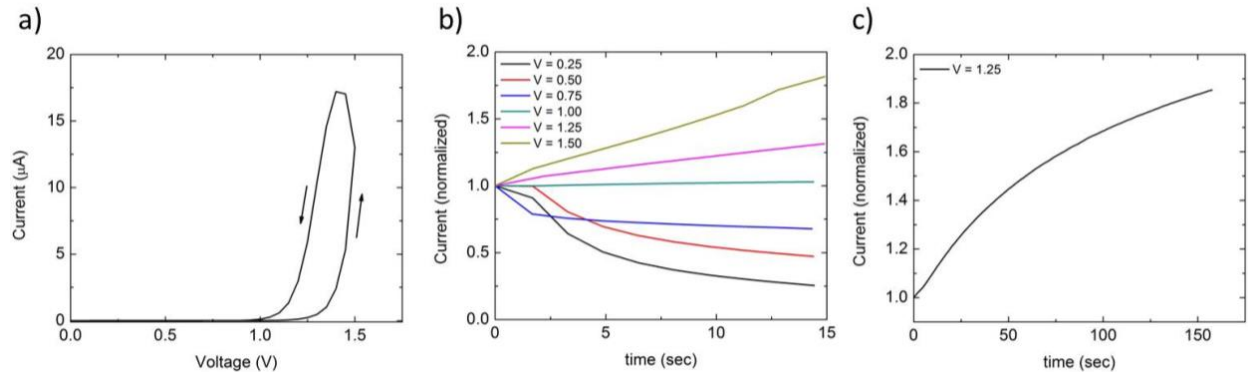

**Figure 1. Current-Voltage and Current-Time characteristics of metal-Nb$_2$O$_{5-x}$-metal memdiodes. a)** Current-Voltage curve for a metal-Nb$_2$O$_{5-x}$-metal device showing rectification and hysteresis. **b)** Current transients at several voltage steps in **(a)** normalized to the first data point showing a decreasing current below the turn-on voltage of 1 volt and increasing current above the turn-on voltage. **c)** An extended current transient above the turn-on voltage showing that the time scale to reach steady state is hundreds of seconds.

In addition to rectification and hysteresis, it is desirable for the memristors in a neuristor circuit to exhibit capacitance. This eliminates the need for an external parallel capacitor, enabling higher density circuits. Capacitance measurements of various sized memdiodes indicate the relative dielectric constant $\varepsilon_r$ is 37.1 ± 1.0 at 1 MHz. As shown in Figure 2a, the memdiode capacitance is dependent on frequency and the capacitance at biological frequencies (< 100 Hz) is significantly higher ($\varepsilon_r$ > 180) than the capacitance at 1 MHz. This native capacitance allows the memdiode to be scaled to nanometer sizes without large external capacitors for application in the neuristor circuit. The device shown in Figure 1 was 400 μm in diameter, however the time required to charge a device (which determines the neuristor oscillation frequency) to reach its switching voltage is not area dependent and was constant for devices with diameters ranging from 100 μm to 1000 μm. This is due to the joint cancelling dependence of current and capacitance on area. While the oscillation frequency of the neuristor is not dependent on the area of the device, it is still dependent on the thickness which provides an alternate method to scale the time required to charge the device.

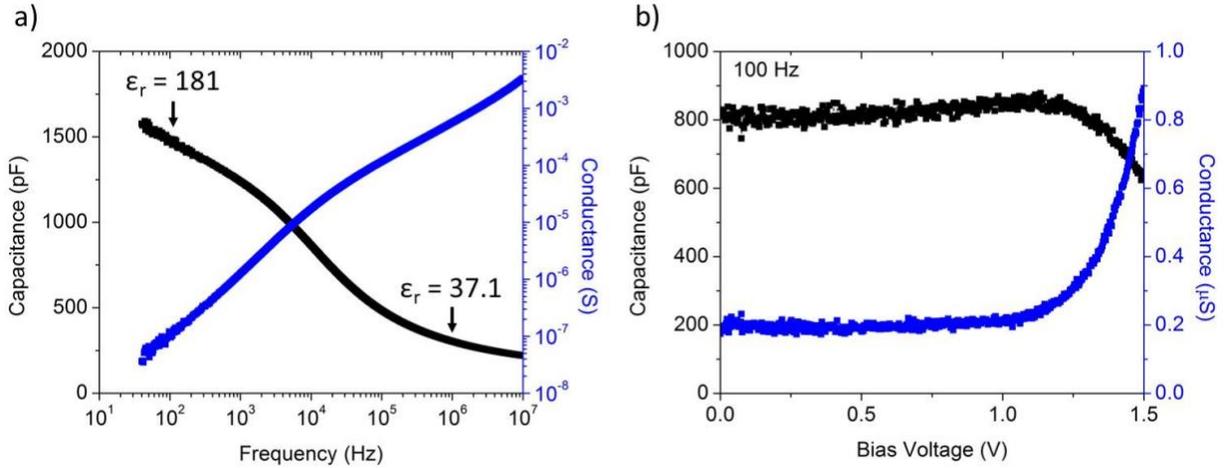

**Figure 2. Capacitance characteristics of metal-Nb$_2$O$_{5-x}$-metal memdiodes. a)** The capacitance and conductance as a function of frequency. The capacitance is highly dependent on frequency with a relative dielectric constant greater than 180 at neurological frequencies (< 100 Hz). **b)** The capacitance and conductance as a function of DC bias voltage measured at 100 Hz. The capacitance is relatively unchanging below the turn-on voltage, increasing by only 6%. The slight increase in capacitance is likely due to image-force lowering decreasing the effective insulator thickness. Above the turn-on voltage the capacitance decreases and the conductance exhibits rectification.

As shown in Figure 2b, the bias dependent capacitance and conductance of the memdiodes shows that, unlike a traditional Schottky or p-n junction diode, the capacitance increases slightly for voltages below the turn-on voltage and decreases above the turn-on voltage. The device conductance exhibits rectification similar to the quasi-steady state measurements shown in Figure 1a. The memdiode therefore maintains the ability to store charge under bias, particularly above turn-on, which is necessary for the neuristor circuit to operate.

Devices of equal thickness (25 nm) were fabricated with Cu, Au, Ni, and Pt contacts in order to investigate the effects of the metal work function on device performance. As shown in Figure 3a, devices exhibit a transition between a high resistance region and a low resistance region when plotted on a semi-log plot. This transition voltage scales linearly with the ideal metal work function[13], as shown in Figure 3b, indicating that the transition is associated with the metal-Nb$_2$O$_{5-x}$ Schottky barrier height. Memdiodes with a minimum transition voltage of 0.128 V were obtained with Cu electrodes, a promising result for low voltage and low power operation. In addition, the electron affinity of amorphous Nb$_2$O$_{5-x}$ is estimated to be 4.5 eV ± 0.07 eV by extrapolating the ideal work function to zero transition voltage.

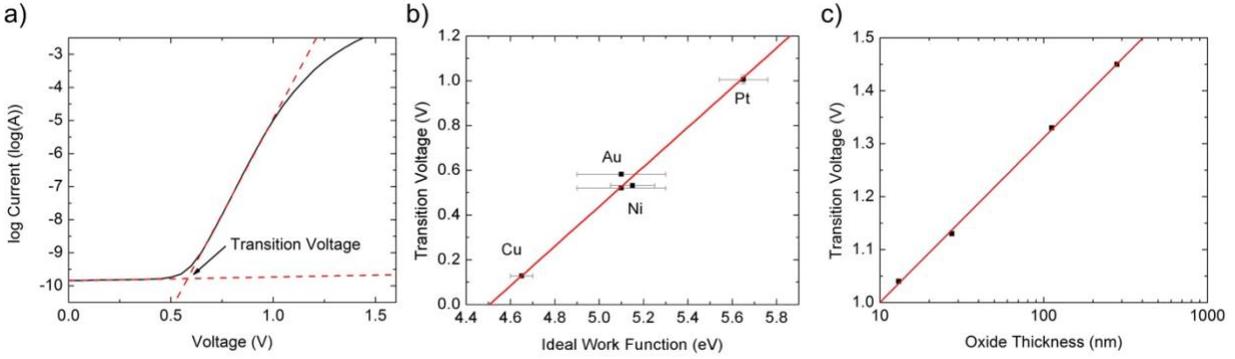

**Figure 3. Linear and logarithmic scaling of the memdiode transition voltage. a)** Semi-log plot of the current-voltage sweep defining a transition voltage between the off and on states. **b)** The transition voltage is linearly proportional to the ideal metal work function indicating that the transition is tied to the Schottky barrier height. **c)** The transition voltage is logarithmically related to the oxide thickness in contrast to traditional MIM diodes.

Devices were also fabricated with a range of $Nb_2O_{5-x}$ thicknesses ranging from 10 nm – 300 nm. As shown in Figure 3c, the transition voltage scales as the logarithm of thickness with thicker devices requiring a larger voltage to transition to the conducting state. This contrasts with a traditional MIM diode based on Fowler-Nordheim tunneling for which the turn-on voltage scales nearly linearly with thickness.

**Discussion**

Rectification in MIM diodes is typically associated with a transition in current mechanism from thermionic emission in the off state to Fowler-Nordheim tunneling in the on state[14]. However, this interpretation should lead to a turn-on voltage which is linearly dependent on the thickness and weakly dependent on the metal-insulator Schottky barrier height. The devices discussed herein exhibit rectification at lower voltages more consistent with the expected Schottky barrier heights for the metal-$Nb_2O_{5-x}$ junction and possess turn-on voltages logarithmically dependent on thickness. This can be explained by considering charge transport via the defect states of the $Nb_2O_{5-x}$ layer and by considering two separately charged regions within the $Nb_2O_{5-x}$ layer. As shown in Figure 4, when the defect states in Region 1 are empty a fixed positive charge and an enhanced electric field result compared to Region 2 where the defects are occupied and thus neutral. This spatially varied charge occupation results in two distinct conduction states. Under low voltage conditions, shown in Figure 4a, the supply of electrons into Region 1 is governed by thermionic emission over a Schottky barrier which limits overall current flow. These injected electrons are swept into Region 2 where they propagate via the Poole-Frenkel mechanism. With each successive voltage increment, the electric field within Region 2 momentarily drives a Poole-Frenkel current larger than the thermionic current depleting electrons from the edge of Region 1. This depletion of electrons leaves behind additional positively charged defects, widening Region 1, which supports a larger proportion of the applied voltage. This conversely reduces the electric field within Region 2 and thus reduces the Poole-Frenkel current to balance the available thermionic current in the steady-state.

After sufficient voltage is applied to reach the device turn-on, shown in Figure 4b, tunneling across the charged Region 1 can occur from the metal electrode directly into Region 2. For niobium oxide with sufficient active defects, the electric field in Region 1 is dominated by the amount of ionized charge in Region 1 causing the turn-on voltage to be controlled by the Schottky barrier height rather than the device thickness as is experimentally observed in Figure 3. As the voltage is increased beyond the turn-on voltage, an increase in the electric field in Region 2 results in an increase in the Poole-Frenkel current. Thus, the Poole-Frenkel current dominates the current flow and dictates the injected tunneling current through control of the width of Region 1 by the emptying (or filling) of defects at the interface between the two regions.

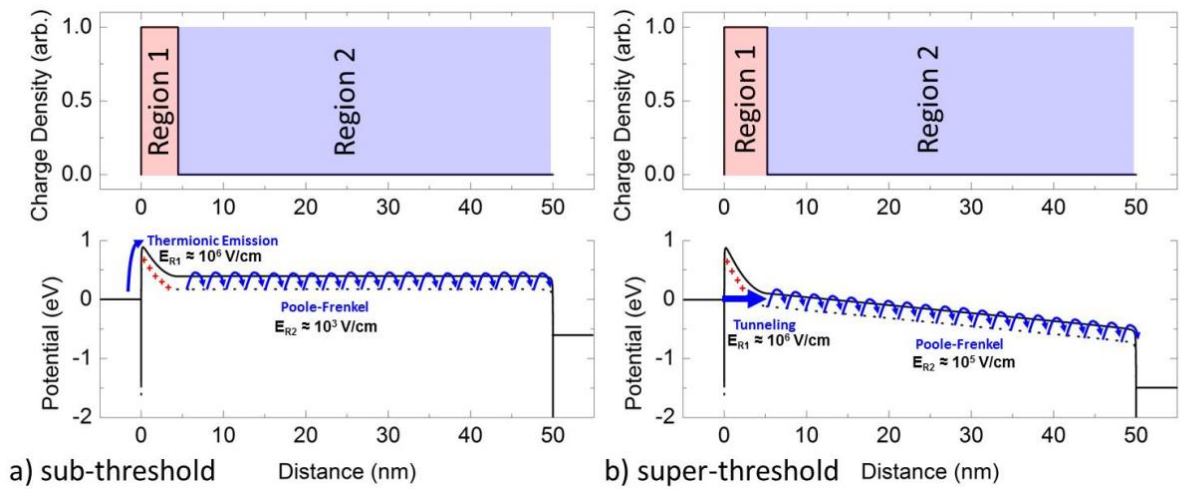

**Figure 4. Charge density maps and band diagrams below and above turn-on.** The charge density and electron potential of the proposed device model both **a)** below and **b)** above the turn-on voltage. Removal of electrons from traps in Region 1 leaves fixed positive charge that supports a large electric field ($10^6$ V/cm). The steady state width of Region 1 is determined by balancing the injection of electrons by thermionic emission below the turn-on voltage or tunneling current above the turn-on voltage with the removal of electrons by Poole-Frenkel current through Region 2. **a)** Below the turn-on voltage a small electric field ($10^3$ V/cm) is required in Region 2 to remove the electrons injected by thermionic emission. **b)** Above the turn-on voltage a larger electric field ($10^5$ V/cm) is required in Region 2 to remove electrons injected by tunneling.

This proposed model for these metal-$Nb_2O_{5-x}$-metal memdiodes predicts that for voltages below turn-on the conduction will be controlled by the Schottky barrier height and the associated thermionic current. Above the turn-on voltage, current will be limited by the Poole-Frenkel current described by equation (1).

$$J_{PF} = q\mu N_o E \exp\left(-\frac{q}{k_B T}\left(\phi_D - \sqrt{\frac{qE}{\pi\varepsilon}}\right)\right) \qquad \text{Equation 1}$$

where $\mu$ is the electron mobility, $N_o$ is the concentration of defects, E is the electric field, $\phi_D$ is the defect trap energy, and $\varepsilon$ is the high frequency dielectric constant[14].

Based on Equation 1, plotting the logarithm of current divided by the electric field, ln(I/E), vs the square root of the electric field will reveal any Poole-Frenkel controlled voltage regimes as a linear region. As shown in Figure 5, for voltages above a threshold value there is a linear regime confirming the Poole-Frenkel current mechanism controls conduction at high voltages.

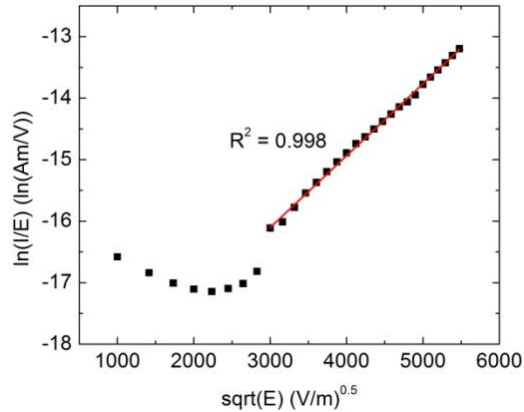

**Figure 5. Poole-Frenkel current characterization.** A Poole-Frenkel plot exhibiting a linear region above the turn-on voltage indicating that the Poole-Frenkel mechanism controls the current flow within this voltage range. The coefficient of determination ($R^2$) is 0.998 indicating good agreement between the experimental measurements and the linear fit.

To determine the trap energy in Equation 1, temperature dependent conductivity measurements were performed at DC biases above the turn-on voltage where the device was shown to be controlled by the Poole-Frenkel current. As shown in Figure 6a, measured at 1.25 V, the temperature dependent current exhibits rectification. The rectification is caused by the supply of electrons transitioning from a tunneling current mechanism at low temperature to thermionic emission current at high temperatures. By fitting $\ln(I/T^2)$ for the thermionic current at high temperatures and $\ln(I)$ for the Poole-Frenkel current, which controls the tunneling current as previously shown, at low temperatures, the Schottky barrier height and the Poole-Frenkel trap depth can be determined.

The Schottky barrier height was determined for the thickest device in Figure 3c, 280 nm, to be 0.96 eV ± 0.08 eV comparable to the theoretical barrier height for Pt/$Nb_2O_5$ of 1.0 eV, but slightly lower than the measured turn-on voltage for this device of 1.12 V. The higher experimental turn-on voltage is likely due to an additional voltage drop over Region 2 in thicker devices as shown in Figure 3c. A less likely explanation for the temperature rectification observed in Figure 6a is activation of a deep defect state that contributes to the Poole-Frenkel current. Fitting the high temperature activation energy of the Poole-Frenkel Arrhenius plot yields a defect energy of 1.0 eV slightly lower than the reported 1.1 eV second ionization energy of an oxygen vacancy in $Nb_2O_5$[6]. Emission from such a deep defect would be extremely slow, resulting in a negligible contribution to current. It is therefore assumed that the observed temperature rectification is due to a transition from a shallow defect Poole-Frenkel current at low temperatures to thermionic emission over the Schottky barrier at high temperatures.

As shown in Equation 1, the Poole-Frenkel trap energy is modified by an electric field term that must be fit by measuring the temperature dependent current at multiple electric fields. As shown in Figure 6b, fitting the measured Poole-Frenkel activation energies yields a defect depth of 0.22 eV ± 0.01 eV, in agreement with the previously reported 0.22 eV activation energy for an oxygen vacancy in amorphous $Nb_2O_5$[6].

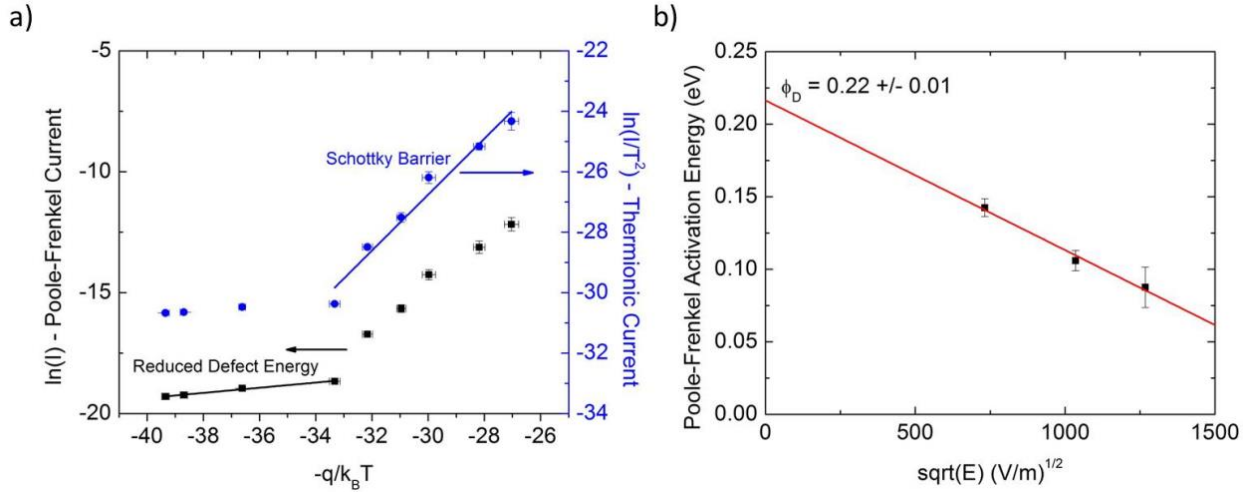

**Figure 6. Temperature dependent Current-Voltage characteristics show a 0.22 eV trap activation energy. a)** Temperature dependent measurements of the device current showing rectification with temperature. The current was measured at 1.25 V, which is above the turn-on voltage. At low temperatures the Poole-Frenkel current controls the current flow with a low activation energy while at high temperatures thermionic emission over a higher energy barrier controls current flow. **b)** Plotting the Poole-Frenkel activation energy at a variety of applied voltages reveals a defect trap depth of 0.22 eV ± 0.01 eV matching the previously reported trap depth for oxygen vacancies in $Nb_2O_5$.

The observed hysteresis, a necessary property for neuristor circuit operation, is caused by non-steady state current transients when operating at sufficiently low frequencies. Low frequency operation, less than 1 kHz, is the primary region of interest for neurological processes[15]. It was experimentally observed that the time scale to reach steady state is on the order of 10's to 100's of seconds indicating any reasonable neuromorphic frequency, f > 0.1 Hz, will exhibit hysteresis. The fixed charge in Region 1 determines the Region 1 width and thus the current through the device. Therefore, the rate of filling and emptying of traps at the boundary between Region 1 and Region 2 sets the time scale to reach a steady state current. Such slow trap dynamics could be caused by either slow emission of electrons from the trap states or a current limited filling and emptying of traps. The capture ($c_e$) and emission ($e_e$) rates for the trap states can be calculated according to Equations 2 – 3[14].

$$c_e = v_{th}\sigma_n n N_t \qquad \text{Equation 2}$$

$$e_e = v_{th}\sigma_n N_C N_t \, exp\left[-\frac{E_C - E_t}{k_B T}\right] \qquad \text{Equation 3}$$

where $v_{th}$ is the thermal velocity, $\sigma_n$ is the electron capture cross section, $n$ is the density of available electrons, $N_t$ is the density of traps, $N_C$ is the effective density of states in the conduction band, and ($E_C - E_t$) is the depth of the trap below the conduction band edge. While the exact values for some of these variables are unknown for amorphous $Nb_2O_5$, typical ranges for these variables are: $v_{th} \approx 10^7$ cm/s, $\sigma_n \in [10^{-14}, 10^{-16}]$ cm$^2$, and $N_C \in [10^{16}, 10^{20}]$ cm$^{-3}$. Based on these values, the timescale for an emission limited process ($N_t/e_e$) is nanoseconds to milliseconds, significantly faster than the experimentally observed current transients. However, for the experimentally observed range of currents, the timescale for a current supply limited capture process ($N_t/c_e$) is milliseconds to kiloseconds, matching the experimental timescale for the current transients. Therefore, the memdiode hysteresis is likely caused by a current limited electron capture process which enables a wide range of frequencies for which the memdiode will exhibit the hysteresis necessary for neuromorphic circuitry.

To corroborate the role of an oxygen deficient active layer in the $Nb_2O_{5-x}$ memdiodes, devices were fabricated at a multiple oxygen stoichiometries for chemical and electrical characterization. Film composition and the formation of oxygen deficiency related subgap states was further investigated using both hard and soft x-ray photoelectron spectroscopy (HAXPES and XPS). The oxygen content was varied by manipulating the Nb target applied power while holding the oxygen flow constant. Three deposition conditions were investigated: 50 W (the standard condition characterized in all other sections of this manuscript), 40 W, and 30 W Nb target power.

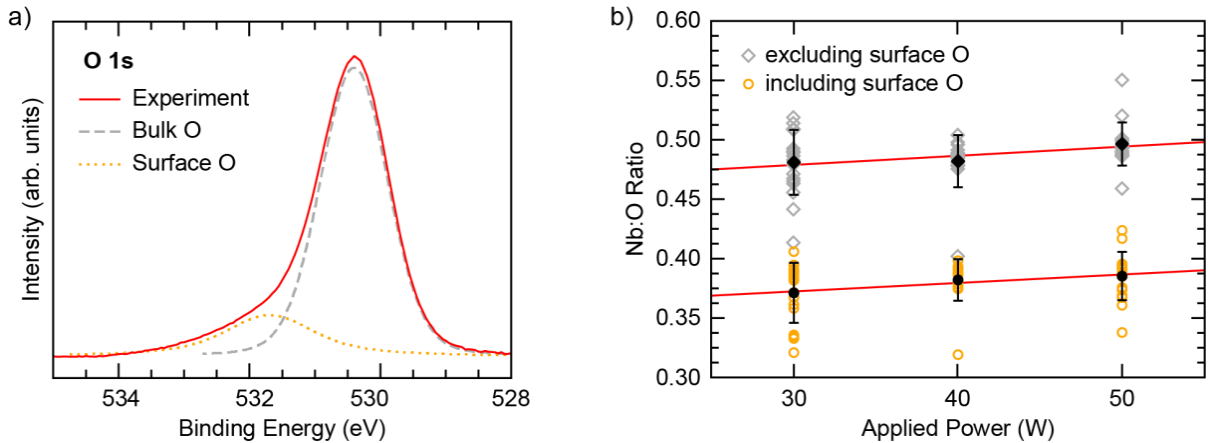

**Figure 7. Average film compositions from XPS.** a) An example O 1s core region from XPS showing the two peaks used during peak fitting as a part of the compositional analysis. b) Nb:O film ratios for 30, 40, and 50 W depositions determined from XPS measurements, both including and excluding surface contamination related oxygen in the ratio.

Figure 7a displays the applied Nb target power dependent average surface compositions for $Nb_2O_{5-x}$ films as determined from XPS core-level measurements of a large sample set (refer to SI for more details). To circumvent issues associated with preferential sputtering during the etching, XPS measurements were performed on films transferred from growth under inert conditions. Multiple growth sets and sample locations were measured to account for chemical variations between films, similar to previous XPS studies of amorphous a-IGZO[16]. While the determined compositions are oxygen rich when including all measured oxygen in the compositional analysis, excluding the oxygen species attributed to surface over-oxidation and contamination from the compositional analysis results in a clear positive trend between oxygen deficiency and applied power during the growth (refer to SI for more details). We report oxygen deficient Nb:O ratios of $0.481 \pm 0.027$, $0.482 \pm 0.021$, $0.497 \pm 0.018$ for 30, 40, and 50 W Nb target powers respectively when excluding surface oxygen from the analysis.

Shown in Figure 7b, these Nb:O ratios can be fit with a straight line, resulting in a slope = $0.0007 \pm 0.0002$, y-intercept = $0.35 \pm 0.01$, and $R^2 = 0.90$ for the ratios including surface oxygen, and a slope = $0.0008 \pm 0.0004$, y-intercept = $0.46 \pm 0.02$, and $R^2 = 0.80$ for the ratios excluding surface oxygen. Regardless of whether the surface over-oxidation is accounted for in the analysis, the same positive trend is observed with nearly identical slopes, confirming the bulk film indeed becomes more oxygen deficient at higher Nb deposition powers.

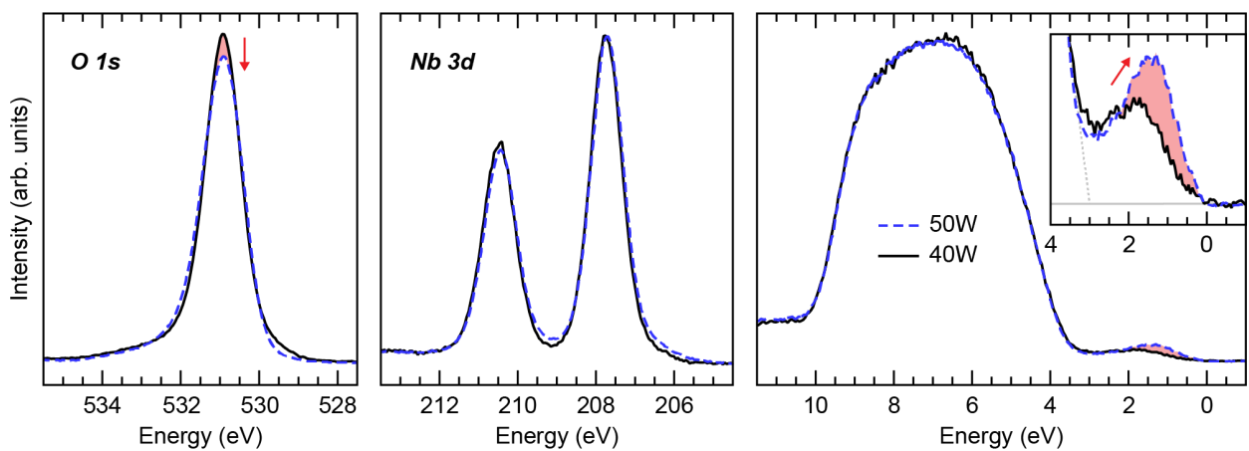

**Figure 8. The role of oxygen stoichiometry in mid-gap state creation.** HAXPES spectra of the core levels and valence band show an increase in mid-gap states at the Fermi Level with decreases in bulk oxygen content. The inset shows the change in mid-gap states, with the estimated valence band maxima (gray dotted line) shown near 3 eV indicating the mid-gap states exist throughout the ~ 3.15 eV band gap.

Hard x-ray photoelectron spectroscopy (HAXPES), which provides a much deeper probing depth than traditional XPS, was performed at beamline I09 at the Diamond Light Source (DLS), Ltd. Although precise compositions cannot be determined from these measurements, the Nb 3d and O 1s core level spectra, shown in Figure 8, confirm that the bulk of the 50 W film possesses a lower oxygen content than the 40 W film after accounting for attenuation caused by the C 1s core level at 248.4 eV and then scaling all core level spectra to normalize the Nb $3p_{3/2}$ peak. In addition, after scaling both valence band spectra to the background, states within the gap above the main valence band edge are found to possess considerably more weight in the 50 W film. These states exist within the ~ 3.15 eV band gap, indicating they are stoichiometry-related defect states or split-off states rather than simple conduction band filling, and are primarily of Nb 4d orbital character. As the Nb:O ratio is increased (with increasing power) the Nb 4d state increases in intensity and shifts towards the expected bulk conduction band minimum (CBM) of pure $Nb_2O_5$. As a result, the 50

W film has more accessible states at 0.22 eV below the CBM than the 40 W film, consistent with the extracted Poole-Frenkel activation energy in Figure 6.

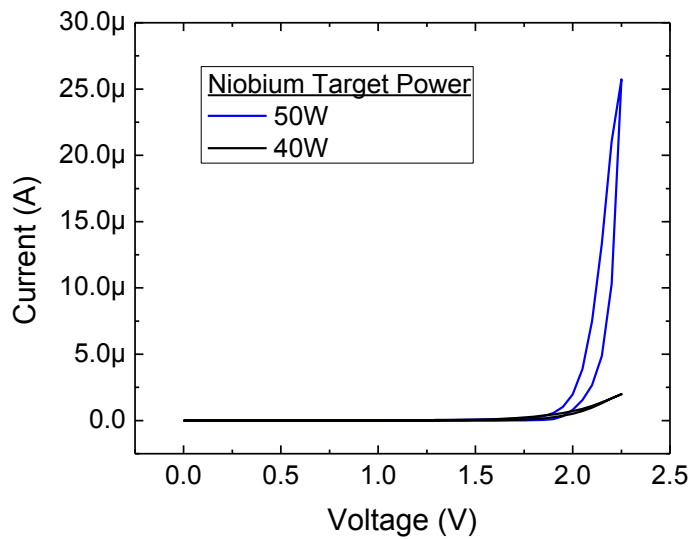

**Figure 9. The role of oxygen stoichiometry in device performance.** Films with higher oxygen stoichiometry turn on in a less defined manner, and the current above turn-on is approximately 10x lower than the lower oxygen stoichiometry films analyzed in the preceding sections.

The current-voltage characteristics of these devices were tested in the same manner shown in Figure 1, and the results are shown in Figure 9. *I-V* data from films with higher oxygen content show an order of magnitude decrease in current at 2.25 V, well above the turn-on voltage. The higher oxygen content devices also show a less defined turn-on, consistent with lower curvature in the band bending across a wider space-charge region due to a lower hole doping profile. This electrical analysis is consistent with the transport model presented above, and when considered with the XPS and HAXPES measurements corroborates the hypothesis that current rectification in metal-$Nb_2O_{5-x}$-metal memdiodes is controlled by Poole-Frenkel defect transport through oxygen deficiencies in the as-deposited material.

**Conclusion**

We have demonstrated a metal-$Nb_2O_{5-x}$-metal device that exhibits the rectification, hysteresis, and capacitance necessary for high density neuristor circuitry without the need to electroform a conducting filament. These devices operate by defect controlled transport. Fixed charge accumulates in oxygen deficiency related defects near the grounded electrode supporting large portions of the applied voltage and creating two regions with differing charge occupations. One region has empty charged defects while the second region has filled neutral defects. Rectification occurs when the voltage drop across the charged region causes the current to transition from thermionic emission in the off state to a Poole-Frenkel current in the on state, which in turn controls tunneling through the charged defect region. The high density of defects causes the turn-on voltage to be dominated by the Schottky barrier height rather than the device thickness in contrast to traditional MIM diodes.

XPS and HAXPES confirm both the deposition condition dependence of oxygen concentration in sputter deposited $Nb_2O_{5-x}$ and the inverse correlation between oxygen concentration and additional subgap states. *I-V* curves corroborate the correlation between oxygen concentration and rectification. Temperature dependent measurements of the Poole-Frenkel activation energy reveal that the defect energy is approximately 0.22 eV ± 0.01 eV in agreement with the presented HAXPES measurements and past measurements of 0.22 eV for oxygen vacancies in amorphous $Nb_2O_5$. Calculating the expected rates at which the current controlling defects can capture and emit electrons reveals that the observed current transients are likely caused by capture rates limited by the supplying current of electrons. This slow trap process limits the speed at which the device can respond creating hysteresis at neurological frequencies.

**Methods**

*Fabrication*

MIM memdiodes were produced on sapphire substrates in three steps. First, a ground plane consisting of a 50 nm titanium adhesion layer and 300 nm of metal (Cu, Au, Ni, or Pt) was deposited by electron beam evaporation across the entire substrate. Next, a layer of niobium oxide was deposited through a shadow mask by room temperature reactive sputtering. Finally, 300 nm thick top metal contacts were deposited through a second shadow mask using electron beam evaporation, matching the underlying metal to form a symmetric device. The top metal contact defines the circular geometry of the devices which range in diameter from 100 μm to 1000 μm.

The niobium oxide insulating layer was sputter deposited at room temperature using a reactive deposition process. A metallic niobium target was sputtered in a mixed $Ar/O_2$ environment using gas flows of 35 sccm and 15 sccm respectively. The chamber pressure was maintained at 10 mTorr. The niobium target was energized with 2.88 $W_{DC}/cm^2$ - 4.80 $W_{DC}/cm^2$ (30 W - 50 W applied power) producing a maximum deposition rate of 5 angstroms per minute. It was previously shown that even low temperature annealing impacts the structural and electrical properties of $Nb_2O_5$[17]. Thus, the low power deposition allowed precise control of the oxide thickness while minimizing unintentional heating of the devices.

*Characterization*

$Nb_2O_{5-x}$ thin films were characterized by x-ray diffraction (XRD), x-ray reflectivity (XRR), optical transmission spectroscopy, x-ray photoelectron spectroscopy (XPS), and hard x-ray photoelectron spectroscopy (HAXPES). Symmetric XRD measurements revealed no peaks except for the sapphire substrate indicating the deposited niobium oxide is amorphous. The density of the deposited niobium oxide was determined by XRR to be 4.1 $g/cm^3$ ± 0.1 $g/cm^3$, significantly lower than the ideal density of 4.55 $g/cm^3$ for $Nb_2O_5$ consistent with oxygen deficient material[17].

Applying the Tauc method to the optical transmission spectra revealed an optical band gap of 3.15 eV ± 0.004 eV, slightly lower than the reported range of band gaps from 3.2 eV – 4 eV[18]. This matches previous reports of a lower band gap for low density $Nb_2O_5$[17]. Additional material characterization details are available in the supplementary information including optßical transmission spectra, XRD, XRR, and XPS. Fabricated memdiodes were electrically tested to determine their steady state current-voltage relation, frequency dependent impedance, and capacitance both with and without bias. Steady state measurements were performed using a Keithley 6517A Electrometer. Frequency dependent measurements were performed using an Agilent 4294A Precision Impedance Analyzer with a frequency range from 40 Hz – 10 MHz.

XPS measurements were performed using a laboratory-based monochromated Al Kα x-ray source with a hemispherical analyzer located at the Analytical and Diagnostics Laboratory at Binghamton University. Measurements were performed with a pass energy of 23.5 eV, corresponding to an instrumental resolution of 0.51 eV, determined from analyzing the Fermi edge and Au $4f_{7/2}$ peak of gold foil references. To reduce surface contamination and oxidation from atmospheric exposure, samples were vacuum sealed after growth before being shipped to Binghamton University. Once at Binghamton University, the samples were opened and mounted for measurement under an inert Ar atmosphere in a glovebox and then transferred to the XPS chamber via a sealed vacuum suitcase.

HAXPES measurements were performed using a photon energy of hv ≈ 5935 eV at beamline I09 of the Diamond Light Source, ltd. synchrotron at the Harwell Science and Innovation Campus in Oxfordshire, UK. The photon beam was monochromated using a channel cut Si (004) crystal, providing an energy resolution of < 0.5 eV. Binding energy calibration was performed using Au foil reference spectra. Due to the reduced surface sensitivity from the much deeper probing depth of HAXPES versus XPS, the samples were not mounted in a glovebox before performing HAXPES measurements, however, they were still vacuum sealed during shipping and stored in a vacuum desiccator when not being mounted or measured.

*Data Availability*

The datasets generated during the current study are available from the corresponding author on reasonable request.

**Acknowledgments**

This work was supported by the Defense Threat Reduction Agency, Basic Research Award # HDTRA-1-12-1-0031 administered by Dr. Jacob Calkins.

This material is based upon work supported by the Air Force Office of Scientific Research under award number FA9550-18-1-0024. We thank Diamond Light Source for access to beamline I09 (SI16005-1), which contributed to the results presented here. We also thank beamline scientists Pardeep Kumar Thakur and Tien-Lin Lee at Diamond Light Source for their assistance.

**Author Contributions**

J.C. conceptualized the devices, recognized the importance of the results, and derived the transport model. J.C. and B.T. prepared the manuscript. A.D. supervised the manuscript and transport model derivation. J.C., B.T., A.W., and B.S. fabricated devices and performed electrical testing. J.C., B.T., and S.H. performed and analyzed XPS data. M.W. and L.P. performed and analyzed HAXPES data. All authors reviewed this manuscript.

**Competing Interests**

The author(s) declare no competing interests.